\documentclass[%
 reprint,
 amsmath,amssymb,
 aps,
]{revtex4-2}

\usepackage{graphicx}
\usepackage{dcolumn}
\usepackage{bm}

\usepackage{todonotes}
\newcounter{todocounter}

\begin{document}

\preprint{APS/123-QED}

\title{Statistical Analysis of Spurious Dot Formation in SiMOS Single Electron Transistors}

\author{Kuan-Chu Chen$^\textup{1}$}
\author{Clement Godfrin$^\textup{1}$}
\author{George Simion$^\textup{1}$}
\author{Imri Fattal$^\textup{1,2}$}
\author{Stefan Kubicek$^\textup{1}$}
\author{Sofie~Beyne$^\textup{1}$}
\author{Bart Raes$^\textup{1}$}
\author{Arne Loenders$^\textup{1,2}$}
\author{Kuo-Hsing Kao$^\textup{3}$}
\author{Danny Wan$^\textup{1}$}
\author{Kristiaan De Greve$^\textup{1,2}$}

\affiliation{
$^\textup{1}$IMEC, Leuven, Belgium
\\$^\textup{2}$
\mbox{Department of Electrical Engineering, KU Leuven,
Leuven, Belgium}
\\$^\textup{3}$
\mbox{Department of Electrical Engineering, National Cheng Kung University, Tainan, Taiwan}
}

\date{\today}

\begin{abstract}
The spatial distribution of spurious dots in SiMOS single-electron transistors (SETs), fabricated on an industrial 300 mm process line, has been statistically analyzed. To have a deeper understanding of the origin of these spurious dots, we analyzed SETs with three different oxide thicknesses: 8 nm, 12 nm and 20 nm. By combining spurious dot triangulation cryo-measurement with simulations of strain, gate bias, and location of the electron wave function, we demonstrate that most spurious dots are formed through the combined effects of strain and gate bias, leading to variations in the conduction band energy. Despite the similar thermal expansion coefficients of polycrystalline silicon gates and single-crystalline silicon substrates, strain remains a crucial factor in spurious dots formation. This learning can be use to optimize the device design and the oxide thickness, to reduce the density of spurious dot while keeping quantum dot tunability. 
\end{abstract}
 
\maketitle

\section{\label{sec:Introduction}Introduction}

Over the past decade, silicon based quantum devices have been fabricated with good single- and two-qubit fidelities \cite{NatNano_2014_Veldhorst, PRApp_2019_Sigillito, NatCom_2022_Noiri, SciAdv_2022, Nature_2022_Xue, NatElec_2022_Zwerver, Nature_2022_Philips, arXiv_2024_Tanttu}, necessary prerequisites for scalable quantum computing hardware. There are two main advantages of silicon-based quantum dots compared to other mature platforms. First, a purified $^{28}$Si layer can provide a nuclear spin-free environment, prolonging the coherence time to more than a microsecond \cite{NatNano_2014_Veldhorst, PRApp_2019_Sigillito, NatCom_2022_Noiri, SciAdv_2022, Nature_2022_Xue, NatElec_2022_Zwerver, Nature_2022_Philips, arXiv_2024_Tanttu}. Second, silicon-based devices are highly compatible with industrial complementary metal-oxide-semiconductor (CMOS) fabrication processes. Silicon-based quantum dots fabricated in 300 mm fabrication process lines have been reported recently \cite{NatCom_2016_Maurand, IEDM_2018_Pillarisetty, NatElec_2022_Zwerver, VLSI_2023_Elsayed}, confirming said compatibility.

Although the number of quantum devices per wafer can be up to $10^4$ \cite{IEDM_2018_Pillarisetty}, the number of successfully coupled qubits in silicon-based quantum circuits remains limited, with only a few demonstrated recently \cite{Nature_2022_Philips, Nature_2024_Neyens}. One reason for this limitation is the existence of undesired quantum dot, so called ``spurious dots". These spurious dots limit the spin qubit system scalability for two main reasons. First, because of their random distribution, they affect the time needed for quantum dot initialization. Typically,  hours of manual tuning are needed to achieve an optimal working point. While fully automated tuning has recently been reported \cite{arXiv_2024_Schuff}, it still takes tens of hours for one qubit to reach the working point. Second, once the device is tuned up, they interact with gate defined quantum dots, affecting their chemical potential and inter-dot tunnel coupling tunability. Spurious dots are universal and appear in qubit devices fabricated in both laboratory \cite{PRB_2009_Nordberg, JAP_2012_Thorbeck} and industrial \cite{VLSI_2022_Niebojewski, ETS_2023_Lorenzelli, IEDM_2022_Kotlyar} environments. Beyond quantum devices, spurious dots are also observed in transistors when cooled to low temperatures \cite{NanoLett_2014_Voisin, EDL_2019_Bonen, EDL_2020_Yang, JEDS_2022_Pati_Tripathi}.

Since spurious dots are widely observed in devices operating in cryogenic environments \cite{PRB_2009_Nordberg, JAP_2012_Thorbeck, VLSI_2022_Niebojewski, ETS_2023_Lorenzelli, IEDM_2022_Kotlyar, NanoLett_2014_Voisin, EDL_2019_Bonen, EDL_2020_Yang, JEDS_2022_Pati_Tripathi}, various mechanisms have been proposed to explain their formation. Spurious dots can be induced by strain \cite{JAP_2012_Thorbeck, AIPAdv_2015_Thorbeck, JAP_2021_Stein, arXiv_2024_Frink}, unintentional doping \cite{NatPhys_2008_Lansbergen}, surface roughness \cite{NanoLett_2014_Voisin}, oxide charges \cite{PRB_2009_Nordberg}, and work function variation\cite{IEEEAccess_2024_Kao}. However, these causes strongly depend on the material, fabrication process, and device design. It is challenging to attribute spurious dots to a specific cause without further analysis.

One method for identifying the cause of spurious dots is to determine their possible locations \cite{JAP_2012_Thorbeck}. To our knowledge, Thorbeck et al. were the first to attempt to locate spurious dots with the assistance of circuit and capacitance simulation \cite{JAP_2012_Thorbeck}. More recently, the electrostatic triangulation technique has been adopted to locate spurious dots using TCAD simulation \cite{SSE_2022_Kriekouki}.

Although methods for locating spurious dots have been proposed, there is still a lack of statistical analysis on their distribution. In this study, we map the locations of spurious dots in eighteen single-electron transistors (SETs) fabricated using an industrial 300 mm process line. SETs are excellent devices for this analysis because they contain a quantum dot, which is the core component of a qubit, and allow for transport measurements. Among these eighteen devices, we studied three different oxide thicknesses: 8 nm, 12 nm, and 20 nm, as the oxide layer is a critical factor in the formation of spurious dots. By combining the locations of spurious dots with simulations, we find that the spurious dots are mainly induced by the combination of strain and gate bias. Additionally, in 8 nm oxide devices, lower gate-stack quality also plays a significant role, as supported by Hall bar measurements.

\section{\label{sec:Experimental}Experimental}

\begin{figure}[!t]
\centerline{\includegraphics[width=1\columnwidth]{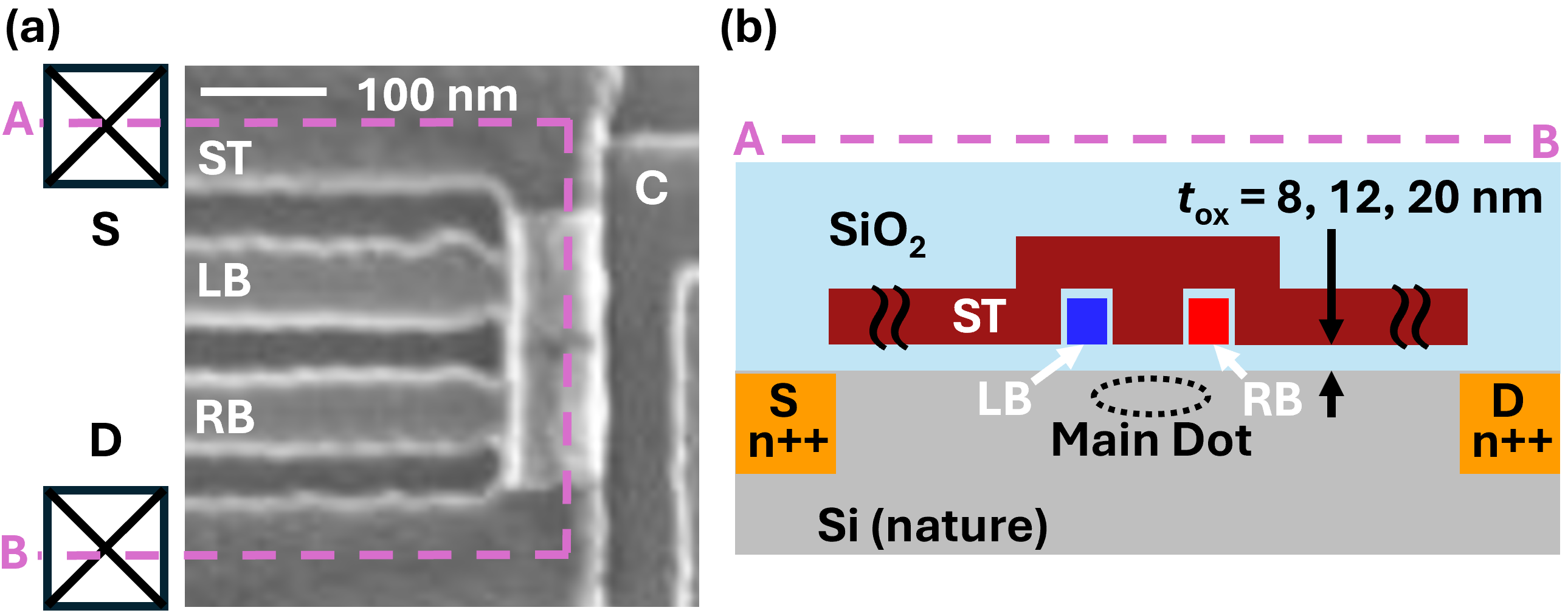}}
\caption{(a) Top-view scanning electron microscope (SEM) image of the SET. (b) Cross-section schematic along the cut line $\overline{\textup{AB}}$ in (a). The source and drain heavily doped regions are located over 10 $\mu$m away from the core part to avoid spurious dots induced by unintentional doping.}
\label{fig1}
\end{figure}

Industrial CMOS fabrication processes are adopted in this work. The fabrication starts with a 300 mm silicon wafer, onto which high-quality oxide layers of 8 nm, 12 nm, and 20 nm are formed by dry thermal oxidation. A three-layer overlapping gate structure is used to build the SET. The gate stacks are formed using heavily doped poly-Si, which is suggested to minimize significant strain \cite{AIPAdv_2015_Thorbeck, IEDM_2019_Mohiyaddin}. The top-view scanning electron microscope (SEM) image and cross-section schematic of the SET are shown in Fig. \ref{fig1}(a) and (b), respectively.
The confinement (C) gate is deposited first and then covered by a 6 nm high-temperature oxide (HTO) layer. The left barrier (LB) and right barrier (RB) gates are deposited as the second gate layer and covered again by a 6 nm HTO layer. Finally, the SET top (ST) gate is deposited as the third gate layer. After the formation of the gate layers, the entire device is covered by a thick layer of chemical vapor deposition (CVD) oxide. Unintentional doping can induce spurious dots \cite{NatPhys_2008_Lansbergen}. Therefore, the source and drain heavily doped regions are placed over 10 $\mu$m away from the core region of the SET. The detailed fabrication process used in this work is also described in Ref. \cite{NpjQuantumInf_2024_Elsayed}.

The devices are measured in a dilution fridge with a base temperature lower than 10 mK and an electron temperature around 135 mK. The source-drain current is measured using standard lock-in detection. During the measurement region of interest, the ST gate is biased at a sufficiently high positive voltage to induce a two-dimensional electron gas (2DEG) between the source and drain, enabling the SETs to turn on. Meanwhile, the C gate is always biased at 0 V to confine the conduction region below the ST gate, preventing it from spreading to the outer parts of the device. The barriers are formed under the LB and RB gates by adjusting their voltages. As shown in Fig. \ref{fig1}(b), when the barriers beneath these gates are properly tuned, the main dot will be formed between them.

\section{\label{sec:Result_and_Discussion}Result and Discussion}

\begin{figure}[!t]
\centerline{\includegraphics[width=1.05\columnwidth]{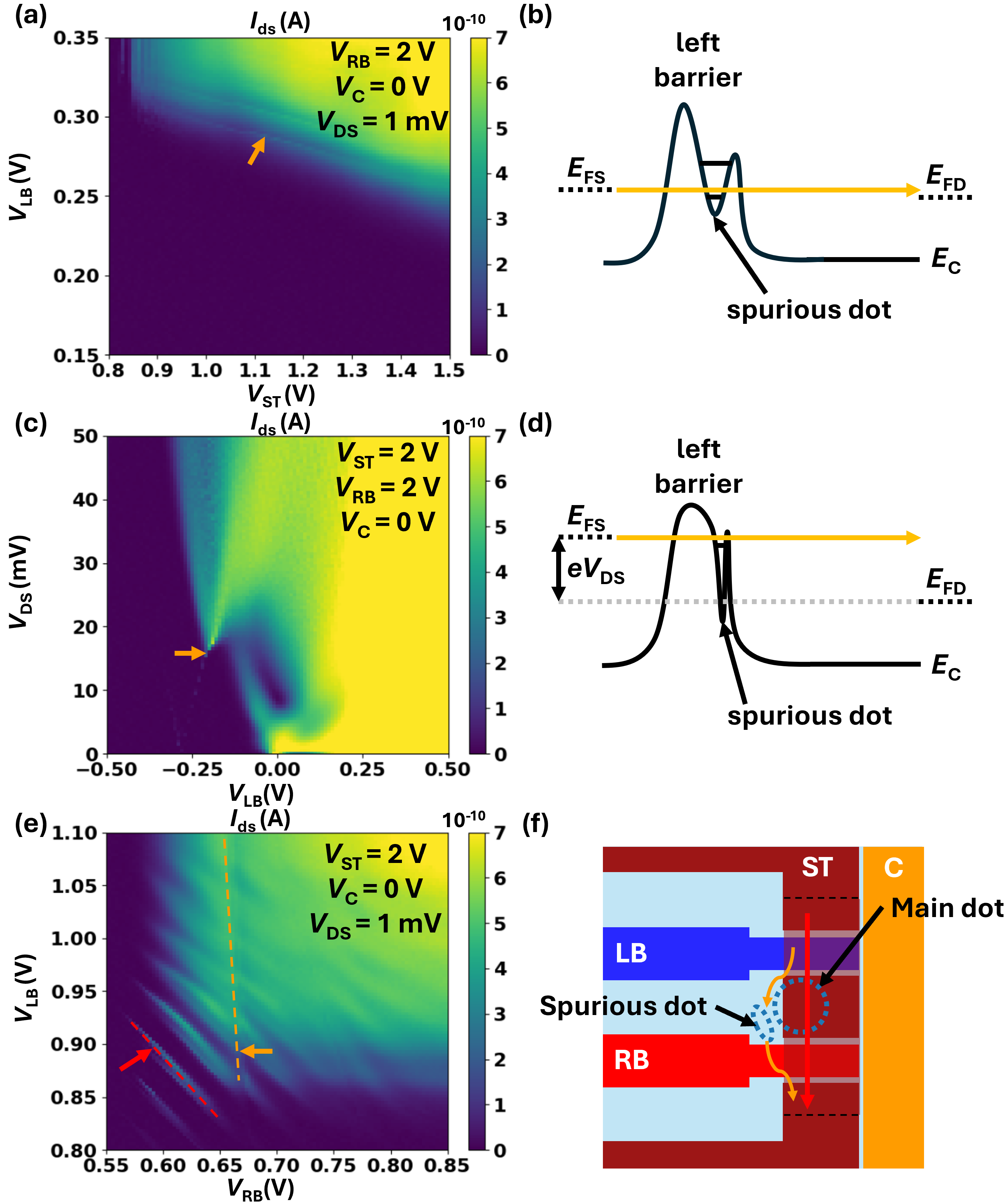}}
\caption{(a) Sweep both $V_\textup{LB}$ and $V_\textup{ST}$ to control the left barrier and observe the spurious dot's signal. (b) Conduction band profile at the bias condition indicated by the orange arrow in (a), where the spurious dot's signal is observed. (c) Sweep both $V_\textup{LB}$ and $V_\textup{DS}$ to observe spurious dots with higher charging energy. (d) Conduction band profile at the bias condition indicated by the orange arrow in (c) when the reservoir's energy is high enough to inject an electron into the spurious dot. (e) Sweep both $V_\textup{LB}$ and $V_\textup{RB}$ to observe of the main quantum dot (red dashed line) and spurious dot (orange dashed line). (f) Schematic of the current path for both the main dot (red arrow) and the spurious dot (orange arrow) observed in (e).}
\label{fig2}
\end{figure}

We have set up a protocol to collect spurious dots statistics over different samples and achieve a comprehensive mapping of their distribution. Five sweeping conditions, listed in Table \ref{table1}, were implemented to observe the spurious dots during the control of the barriers and the main dot.
\begin{table}[!t]
\caption{\label{table1}%
Conditions to observe spurious dots}
\begin{ruledtabular}
\begin{tabular}{cccccc}
\multicolumn{1}{c}{\textrm{$V_\textup{ST}$ (V)}}&
\multicolumn{1}{c}{\textrm{$V_\textup{LB}$ (V)}}&
\multicolumn{1}{c}{\textrm{$V_\textup{RB}$ (V)}}&
\multicolumn{1}{c}{\textrm{$V_\textup{C}$ (V)}}&
\multicolumn{1}{c}{\textrm{$V_\textup{DS}$ (mV)}}&
\multicolumn{1}{c}{\textrm{}}\\
\hline
0 $\sim$ 2 & -0.5 $\sim$ 2 & 2      & 0      & 1    & LB control \\
0 $\sim$ 2 & 2     & -0.5 $\sim$ 2  & 0      & 1    & RB control \\
2     & -0.5 $\sim$ 2 & 2      & 0      & 1 $\sim$ 50 & LB control\textsuperscript{a} \\
2     & 2     & -0.5 $\sim$ 2  & 0      & 1 $\sim$ 50 & RB control\textsuperscript{a} \\
2     & -0.5 $\sim$ 2 & -0.5 $\sim$ 2  & 0      & 1    & dot control \\
\hline
\end{tabular}
\end{ruledtabular}
\footnotetext[1]{Barrier control with different reservoir's energy.}
\end{table}
The measured results and schematic of these spurious dots are shown in Fig. \ref{fig2}.
\begin{figure*}[!t]
\centerline{\includegraphics[width=0.95\textwidth]{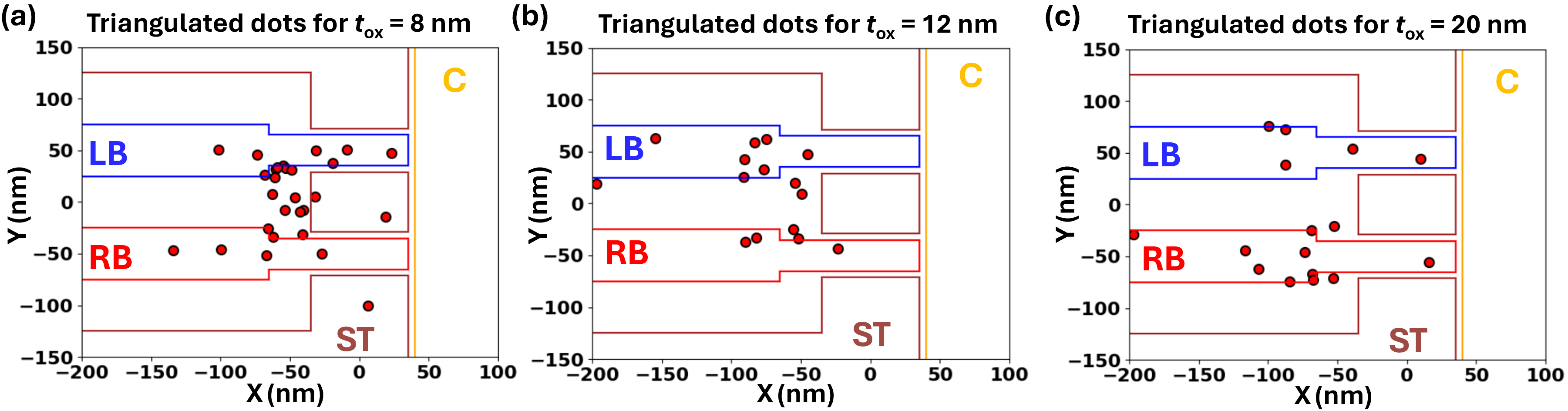}}
\caption{Distribution of spurious dots in devices with oxide thicknesses of (a) 8 nm, (b) 12 nm, and (c) 20 nm. Spurious dots are primarily located at the corners and edges of barrier gates. In 8 nm oxide devices, some spurious dots are also randomly distributed beyond these regions.}
\label{fig3}
\end{figure*}
\begin{figure}[!t]
\centerline{\includegraphics[width=0.62\columnwidth]{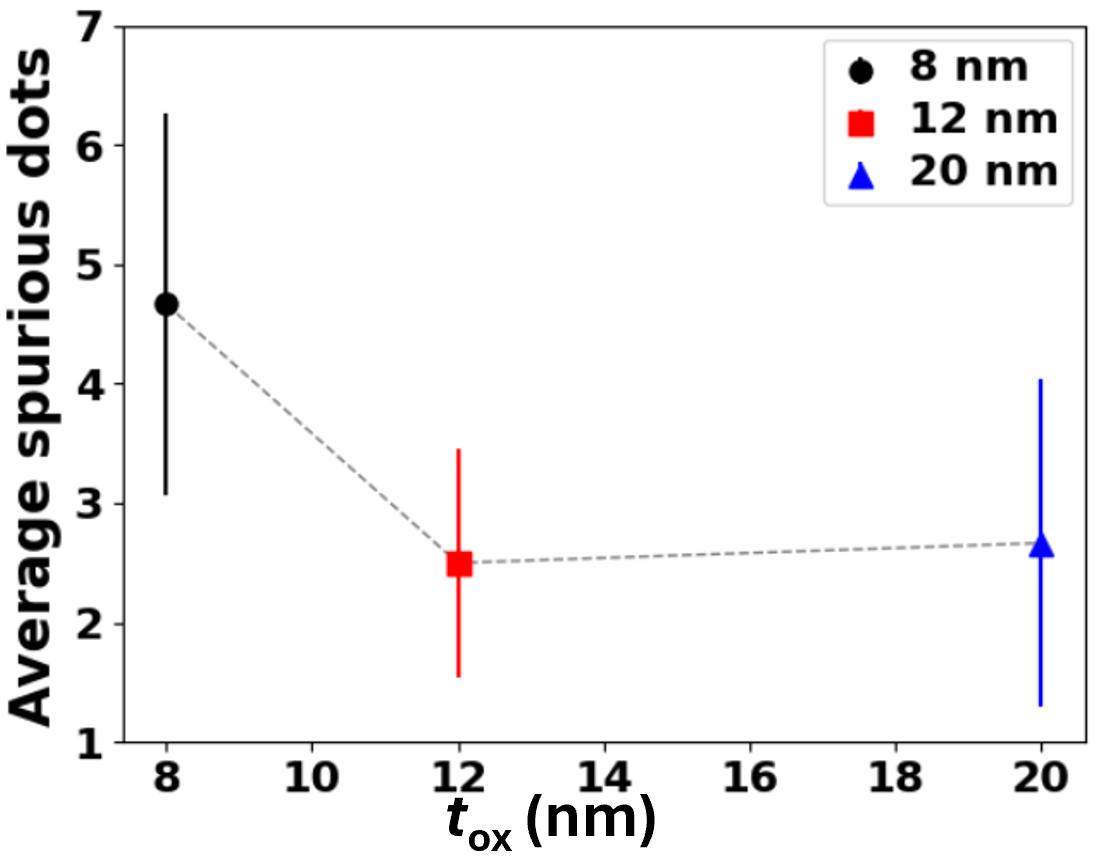}}
\caption{The average number of spurious dots per device for different oxide thicknesses, with error bars representing the standard deviation.}
\label{fig4}
\end{figure}
\begin{figure}[!t]
\centerline{\includegraphics[width=0.62\columnwidth]{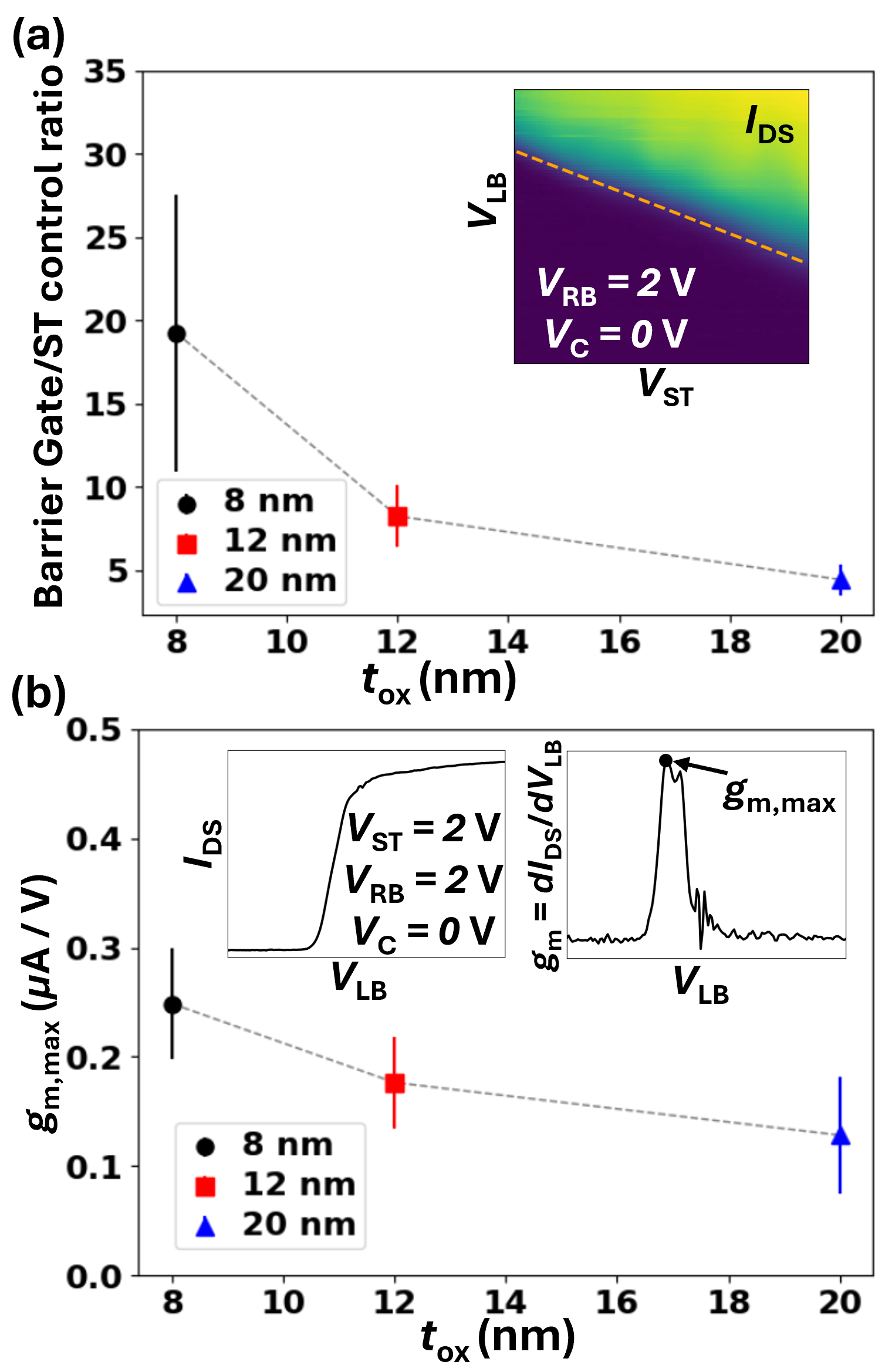}}
\caption{(a) Barrier and (b) channel control of thirty six barrier gates with three different oxide thicknesses. The extraction method for barrier and channel control is indicated in the insets of (a) and (b), respectively. As expected, both barrier and channel control decrease with increasing oxide thickness.}
\label{fig5}
\end{figure}
In Fig. \ref{fig2}(a), $V_\textup{RB}$ is biased at 2 V, forming only the left barrier. In this measurement, the height of the left barrier is controlled by both $V_\textup{LB}$ and $V_\textup{ST}$. As shown in Fig. \ref{fig2}(b), if the combination of $V_\textup{LB}$ and $V_\textup{ST}$ aligns the available energy level of a spurious dot with the energy of the source and drain reservoirs, the transport current will exhibit the form of Coulomb oscillations, as indicated by the orange arrow in Fig. \ref{fig2}(a). Some spurious dots with higher charging energy are difficult to observe during low $V_\textup{DS}$ bias. This scenario is shown in Fig. \ref{fig2}(c) and (d). In Fig. \ref{fig2}(c), we sweep $V_\textup{LB}$ and $V_\textup{DS}$ while forming only the left barrier. As plotted in Fig. \ref{fig2}(d), when $V_\textup{DS}$ is high enough to inject electrons into the available energy level of the spurious dot, the coulomb oscillations will appear, as pointed out by the orange arrow in Fig. \ref{fig2}(c). Finally, we properly control LB and RB to form the main dot and observe the spurious dot during our manipulation. Fig. \ref{fig2}(e) shows the measured result: a series of clear diagonal lines indicating the current contribution from the main dot, as guided by the red dashed line. However, under certain bias conditions, another line, guided by the orange dashed line, may be attributed to a nearby spurious dot. The schematic of this nearby spurious dot is shown in Fig. \ref{fig2}(f).
\begin{figure*}[!t]
\centerline{\includegraphics[width=1\textwidth]{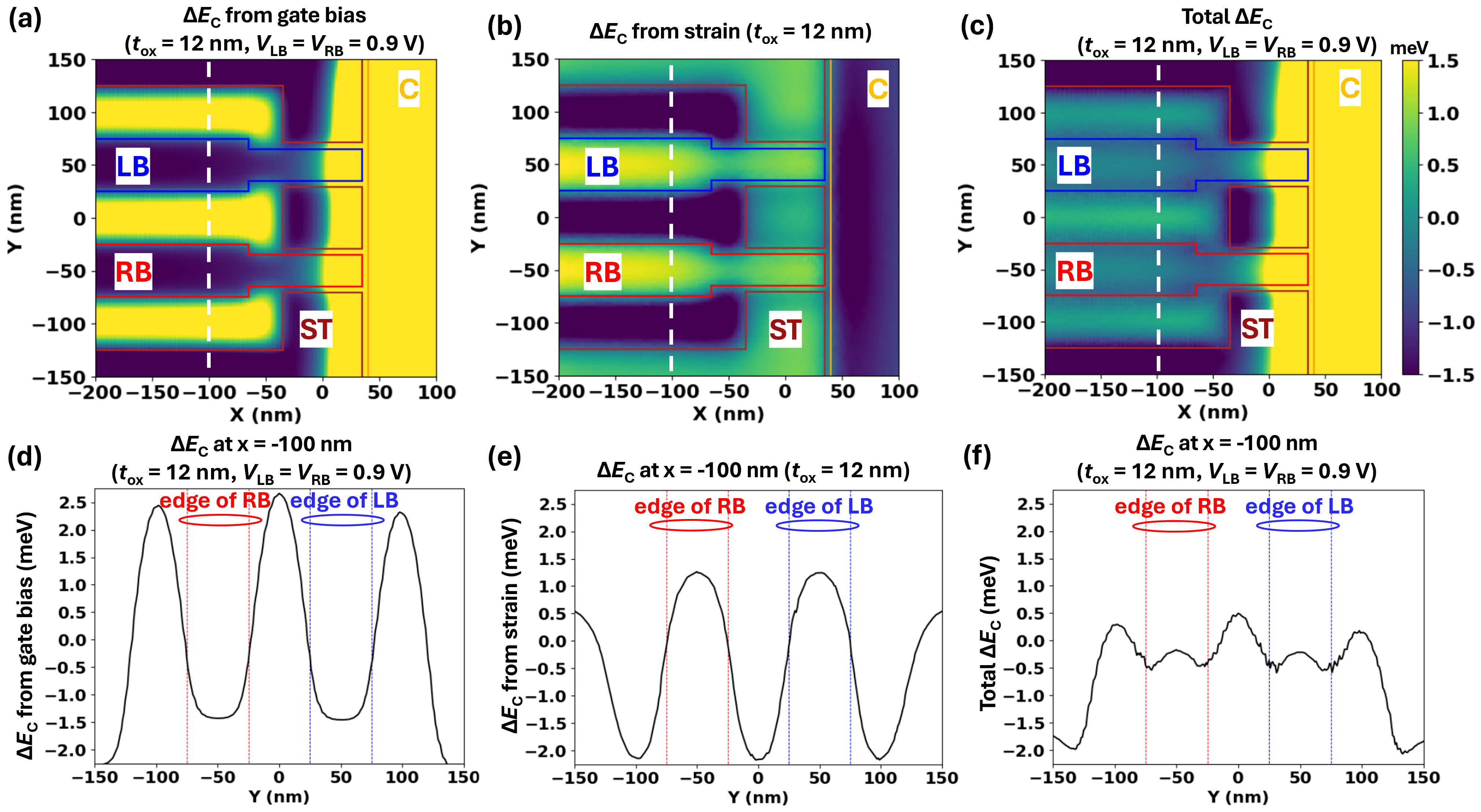}}
\caption{Conduction band energy variation $\Delta E_\textup{C}$ at 1 nm below $\textup{SiO}_\textup{2}/\textup{Si}$ interface, induced by (a) gate bias, (b) strain, and (c) the combination of both for $t_\textup{ox} = 12\textup{ nm}$, $V_\textup{LB} = V_\textup{RB} = 0.9\textup{ V}$. The combined effects of gate bias and strain result in spurious dots forming at the corners and edges of barrier gates, consistent with the experimental observations in Fig. \ref{fig3}. To clarify the behavior of $\Delta E_\textup{C}$, cut lines at $\textup{x = -100 nm}$ from (a), (b), and (c) are illustrated in (d), (e), and (f), respectively.}
\label{fig6}
\end{figure*}

After identifying the bias conditions under which the spurious dots emerge, electrostatic triangulation is used to determine the locations of the spurious dots. The validation and detailed procedure of the triangulation are provided in Appendix \ref{appendix}. The locations of spurious dots for oxide thicknesses of 8 nm, 12 nm, and 20 nm are shown in Fig. \ref{fig3}(a), (b), and (c), respectively. To gain a broader understanding of the spurious dots distribution, locations from six different devices are overlapped for each oxide thickness. Spurious dots are commonly found at the corners and edges of barrier gates. For the 8 nm oxide devices, some spurious dots are observed to be randomly distributed beyond the corners and edges. Additionally, Fig. \ref{fig4} shows the average number of spurious dots for each oxide thickness, with error bars indicating the standard deviation. The highest number of spurious dots is found in devices with the 8nm oxide layer.

Barrier and channel control have been further extracted. A total of thirty-six barrier gates were investigated, with twelve for each oxide thickness. The variation of barrier and channel control with respect to oxide thickness is shown in Fig. \ref{fig5}. The extraction method is illustrated in the inset figures of Fig. \ref{fig5}(a) and (b), for barrier and channel control, respectively.

We determine the control ratio between barrier gates and the ST gate by extracting the slope of the points with the same current level in $V_\textup{LB}, V_\textup{ST}$ and $V_\textup{RB}, V_\textup{ST}$ sweepings, as the points with the same current level indicate the same barrier height \cite{SSE_2022_Kriekouki}. The channel control is estimated by the maximum transconductance $g_\textup{m,max}$ of the channel under the barrier gates. From Fig. \ref{fig5}, it is not surprising that the barrier and channel control both decrease with increasing oxide thickness. This result emphasizes that for quantum devices, the selection of oxide thickness is a critical consideration. It is necessary to have a thick enough oxide to avoid spurious dots while still being thin enough to maintain gate control.

With the locations of the spurious dots demonstrated in Fig. \ref{fig3}, simulations are further conducted to identify the mechanisms leading to their formation. Fig. \ref{fig6}(a) and (b) illustrate the simulated conduction band energy variation $\Delta E_\textup{C}$ at 1 nm below $\textup{SiO}_\textup{2}/\textup{Si}$ interface, induced by gate bias and strain, respectively. In Fig. \ref{fig6}(a), Synopsys TCAD is used to simulate the gate bias induced $\Delta E_\textup{C}$ at $T\textup{ = 10 K}$. For the strain simulation in Fig. \ref{fig6}(b), the Mathematica is employed to calculate strain-induced $\Delta E_\textup{C}$ at $T\textup{ = 0 K}$. Given that the thermal expansion coefficient (TEC) of polysilicon ($\alpha _\textup{poly-Si}$) \cite{JAP_2000_Tada} is typically $0\sim 70\%$ higher than that of single crystalline silicon ($\alpha _\textup{Si}$) \cite{IntJThermo_2004_Watanabe}, depending on temperature, we assume $\alpha _\textup{poly-Si}/\alpha _\textup{Si} = 1.4$ across all temperatures in our simulation. Moreover, the system is relaxed to zero strain at 1200 K, which is the annealing temperature for forming the polysilicon gate. The result in Fig. \ref{fig6}(b) from the strain simulation indicates that although polysilicon is often recommended as a gate material to relax strain \cite{AIPAdv_2015_Thorbeck, IEDM_2019_Mohiyaddin}, it can still induce a conduction band variation $\Delta E_\textup{C}$ on the order of few meV at $T\textup{ = 0 K}$.

\begin{figure*}[!t]
\centerline{\includegraphics[width=1\textwidth]{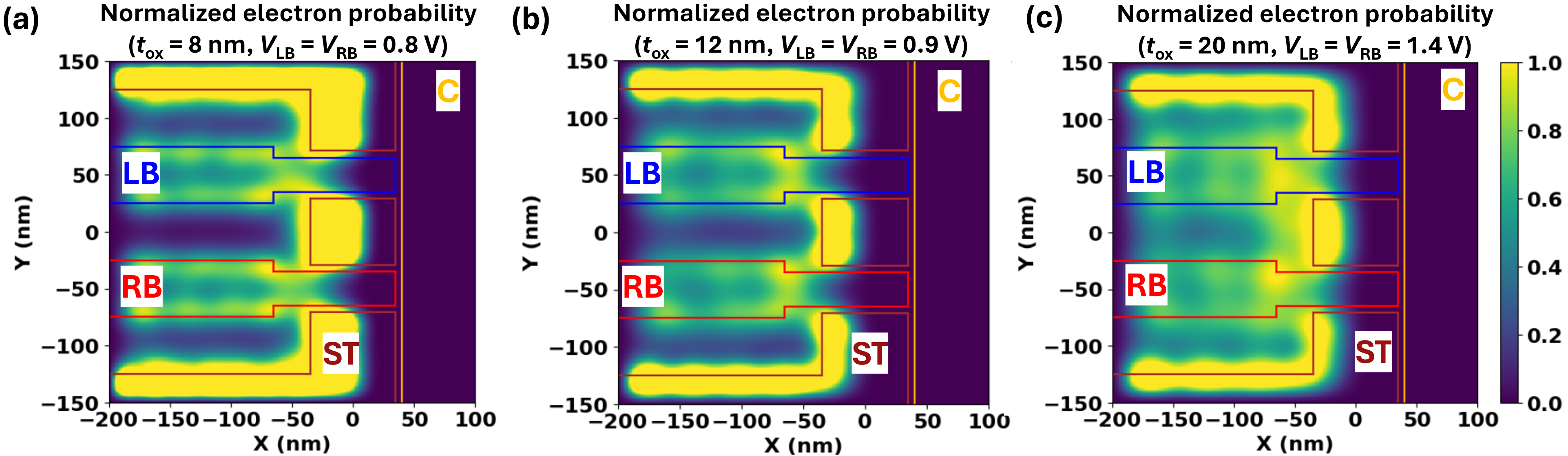}}
\caption{The normalized probability of finding an electron based on the $\Delta E_\textup{C}$ profile induced by the combined effects of gate bias and strain for (a) $t_\textup{ox} = 8\textup{ nm}$, $V_\textup{LB} = V_\textup{RB} = 0.8\textup{ V}$, (b) $t_\textup{ox} = 12\textup{ nm}$, $V_\textup{LB} = V_\textup{RB} = 0.9\textup{ V}$, and (c) $t_\textup{ox} = 20\textup{ nm}$, $V_\textup{LB} = V_\textup{RB} = 1.4\textup{ V}$. The probability values are normalized to the local maximum at the corners of the barrier gates.}
\label{fig7}
\end{figure*}
However, the simulated results in Fig. \ref{fig6}(a) and (b) do not support the spurious dots locations identified in Fig. \ref{fig3}. To further clarify, cut lines of $\Delta E_\textup{C}$ at $\textup{x = -100 nm}$ from Fig. \ref{fig6}(a) and (b) are plotted in Fig. \ref{fig6}(d) and (e), respectively. Fig. \ref{fig6}(d) shows that gate bias tends to induce electron accumulation under the barrier gates. On the other hand, Fig. \ref{fig6}(e) suggests that electrons are more likely to appear between the barrier gates. Neither scenario supports the formation of spurious dots at the corners and edges of the barrier gates, as observed experimentally.

Fig. \ref{fig6}(c) represents the summation of the conduction band variation from Fig. \ref{fig6}(a) and Fig. \ref{fig6}(b), where trenches of $\Delta E_\textup{C}$ are observed at the corners and edges of the barrier gates. The cut line of Fig. \ref{fig6}(c) at $\textup{x = -100 nm}$, shown in Fig. \ref{fig6}(f), reveals distinct valleys in $\Delta E_\textup{C}$, appearing at the edges of the barrier gates. This analysis indicates that spurious dots form through the combined effects of gate bias and strain. Furthermore, this explains why some spurious dots are only observed under specific bias conditions.

To further clarify the formation of spurious dots in the devices, we solve the Schrödinger equation for the conduction band variation induced by the combined effects of gate bias and strain. After solving the Schrödinger equation, the probability of finding electrons in all states with energy lower than the local maximum of $\Delta E_\textup{C}$ between the two barrier gates is summed up. The calculated results at specific bias voltages for three different oxide thicknesses are illustrated in Fig. \ref{fig7}. To enhance the visibility of electron probabilities under the barrier gates, these values are normalized to the local maximum at the corners of the barrier gates. From Fig. \ref{fig7}, it is evident that, apart from the conduction path under the ST gate, electrons have a higher probability of being found at the edges and corners of the barrier gates, which is consistent with the triangulated spurious dots observed in Fig. \ref{fig3}.

\begin{figure}[!t]
\centerline{\includegraphics[width=0.7\columnwidth]{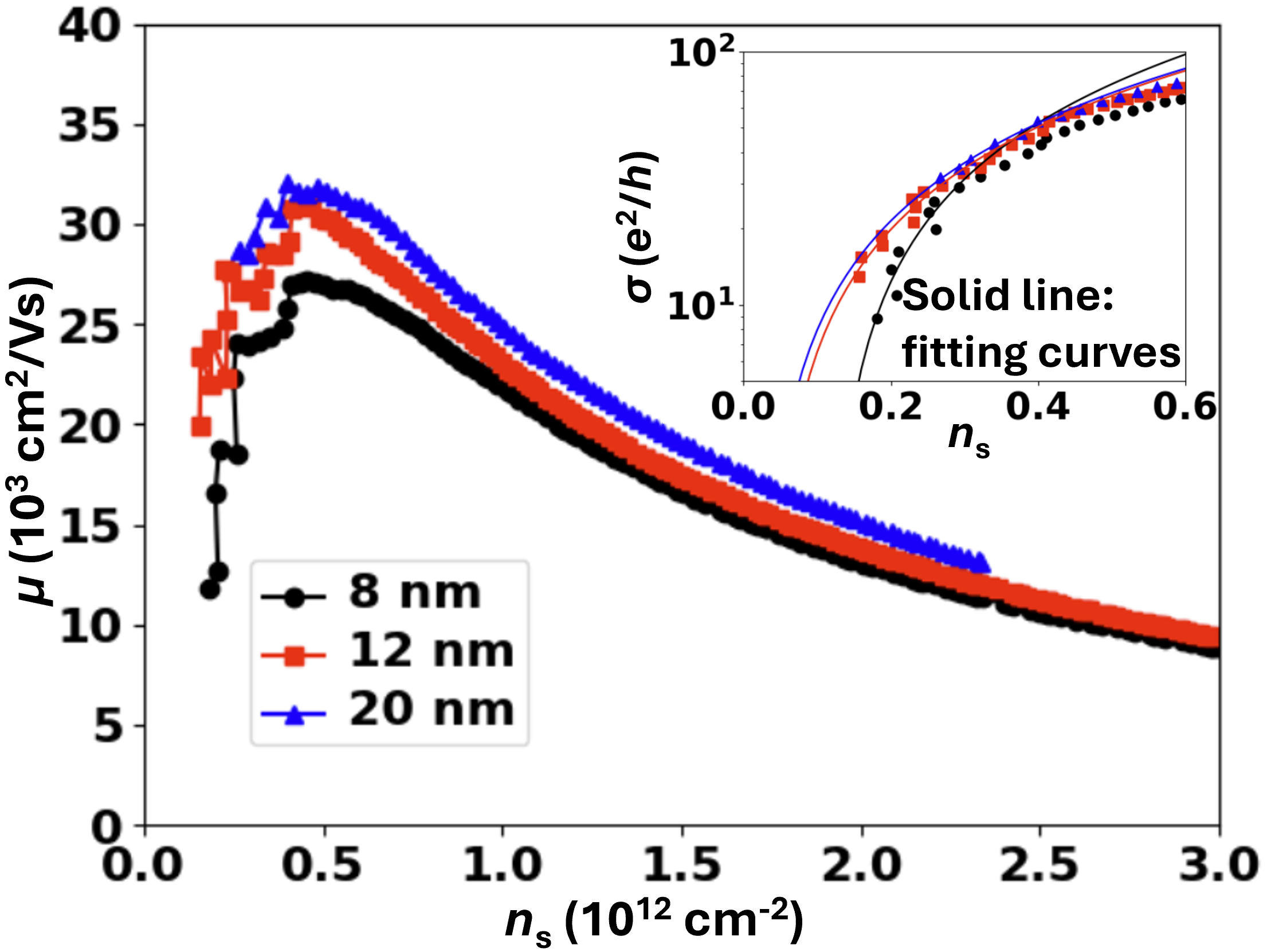}}
\caption{Extracted $\mu$ v.s. $n_\textup{s}$ at $T = 100\textup{ mK}$ from Hall bars with three different oxide thicknesses. The inset shows the sheet conductivity $\sigma$  and the solid fitting lines by using $\sigma = (n_\textup{s} - n_\textup{p})^p$ at $p = 1.2$. The fitted percolation densities $n_\textup{p}$ are 1.19, 0.38, and 0.26 $\times\textup{10}^\textup{11} \textup{cm}^\textup{-2}$ for 8 nm, 12 nm, and 20 nm gate oxide devices, respectively.}
\label{fig8}
\end{figure}

However, in Fig. \ref{fig3}(a), several spurious dots are randomly distributed between the barrier gates and under the ST gate for gate oxide 8 nm devices. It is suggested that the charge-rich interface between the poly-Si gate stack and $\textup{SiO}_\textup{2}$ is closer to the $\textup{Si/SiO}_\textup{2}$ interface, leading to the formation of spurious dots through charge clustering \cite{PRB_2009_Nordberg}. To investigate the impact of remote charge scattering, Hall bar measurements were performed for the three different oxide thicknesses. The Hall bars, with dimensions of $W = 5 \mu\textup{m}$ and $L = 50 \mu\textup{m}$, are designed with a larger size, meaning that strain effects from fine structures are no longer present. The electron mobility $\mu$ and sheet conductivity $\sigma$ at $T = 100\textup{ mK}$ are extracted, as shown in Fig. \ref{fig8}.
By fitting $\sigma$ with the equation $\sigma = (n_\textup{s} - n_\textup{p})^p$ at $p = 1.2$, the highest percolation density was observed in the 8 nm oxide Hall bar, indicating more charge scattering \cite{PRB_2009_Tracy}.

\section{\label{sec:Conclusion}Conclusion}

In this study, we performed a statistical analysis of spurious dots formation in 18 SiMOS single-electron transistors (SETs) fabricated using an industrial 300 mm process line with varying oxide thicknesses of 8 nm, 12 nm, and 20 nm. The results show that SETs with 8 nm gate oxide exhibit approximately twice as many spurious dots compared to devices with thicker oxides. By combining electrostatic triangulation with simulations of gate bias, strain and location of the electron wave function, we found that most spurious dots are localized at the corners and edges of the barrier gates, driven primarily by the interplay between gate bias and strain. Additionally, Hall bar measurements suggest that oxide charges at poly-Si gate stack and $\textup{SiO}_\textup{2}$ interface play a key role in the increased number of spurious dots in 8 nm oxide devices. The electrostatic triangulation method has been proven to be an effective way to identify the mechanisms behind spurious dots.

In the device design aspect, this work highlights that even with a poly-Si gate, which has a TEC similar to that of single-crystalline silicon and is designed to relax strain, it can still cause variations in conduction band energy on the order of a few meV, contributing to spurious dots formation. While increasing oxide thickness reduces the number of spurious dots, it compromises the gate control ability. Therefore, selecting an optimal oxide thickness is critical for minimizing spurious dots while maintaining sufficient gate control in quantum devices.

\begin{acknowledgments}
This work was supported in part by European Union’s Horizon 2020 Research and Innovation Program under grant agreement No 951852 (QLSI), and in part by the National Science and Technology Council (NSTC), Taiwan under contract 113-2917-I-564-007. This work was performed as part of IMEC’s Industrial Affiliation Program (IIAP) on Quantum Computing.
\end{acknowledgments}

\appendix

\section{\label{appendix}Validity of Electrostatic Triangulation}

\renewcommand{\thefigure}{A\arabic{figure}}
\setcounter{figure}{0}

\begin{figure}[!t]
\centerline{\includegraphics[width=1.05\columnwidth]{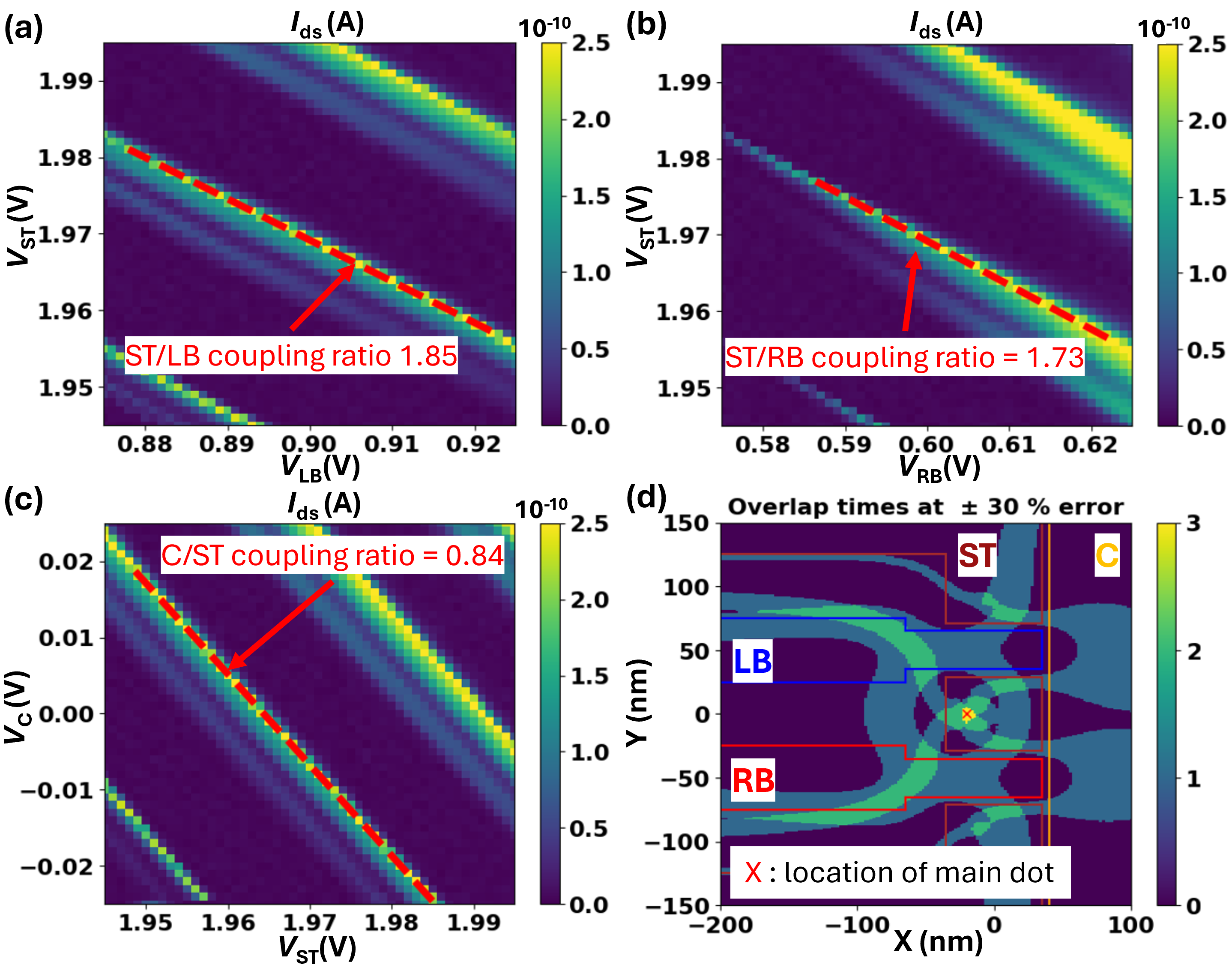}}
\caption{Bias sweeps on (a) $V_\textup{LB}$ and $V_\textup{ST}$, (b) $V_\textup{RB}$ and $V_\textup{ST}$, and (c) $V_\textup{C}$ and $V_\textup{ST}$, around the bias point indicated by the red arrow in Fig. \ref{fig2}(e), to extract the coupling ratio of the main dot between different gates. (d) Overlapping of possible regions, based on coupling ratios, demonstrates that the triangulation of the main dot aligns with the position predicted by 3D TCAD simulations. The outlines of each gate are also included for reference.}
\label{figA1}
\end{figure}
To validate the triangulation method, we apply it to the main dot, whose location is already well known. Our focus is on the bias voltages near the CBO of the main dot. Fig. \ref{figA1}(a), (b), and (c) illustrate the coupling ratios between various gates for the main dot around the bias point marked by the red arrow in Fig. \ref{fig2}(e).
In Fig. \ref{figA1}(a) to (c), the bias voltages on two gates are swept to generate 2D mappings, where the slope of the CBO (Coulomb Blockade Oscillation) lines represents the coupling strength ratio between the two gates. To further analyze these coupling ratios, 3D TCAD electrostatic simulations at $T\textup{ = 10 K}$ were employed to map the possible regions of coupling. The gate dimensions shown in Fig. \ref{fig1}(b) were used for the simulations, and a $\pm 30\%$ margin of error was applied to define the possible regions, consistent with \cite{SSE_2022_Kriekouki}. For the 8 nm gate oxide devices, a margin of $\pm 60\%$ error was adopted to achieve better overlap, as the coupling ratios change more sharply in thin gate oxide devices. The highest possible location of the dot was then determined by overlapping all possible regions, as depicted in Fig. \ref{figA1}(d), which also includes the outlines of the gates. The dot location, highlighted in yellow in Fig. \ref{figA1}(d), aligns precisely with the red cross, which marks the simulated central point of the main dot. In Fig. \ref{fig3}, the center of mass of the overlapping region is identified as the position of the spurious dots. This triangulation confirms the accuracy and validity of our method for locating the spurious dots.

\bibliography{apssamp}

\end{document}